\documentclass[12pt,aps,prb,superscriptaddress,reprint]{revtex4-1}

\usepackage{lmodern}
\usepackage{float}
\usepackage{microtype}
\usepackage{xspace}
\usepackage{amsmath,amssymb,amsfonts,amsthm,mathtools}
\usepackage{mathrsfs,bm}
\usepackage[colorlinks=true, linkcolor = blue, hyperfootnotes=false, citecolor = blue]{hyperref}
\usepackage[T1]{fontenc}
\usepackage[latin1]{inputenc}
\usepackage[english]{babel}
\usepackage{graphicx}
\newcommand{\norm}[1]{\left\lVert #1 \right\rVert}
\newcommand{\av}[1]{\left\langle{#1}\right\rangle}
\newcommand{\dd}{\,\mathrm{d}}
\renewcommand{\refeq}[1]{(\ref{#1})}
\renewcommand{\t}[1]{\bm{#1}}
\renewcommand\[{\begin{equation}}
\renewcommand\]{\end{equation}}

\begin{document}
\title{Elastic quantum spin-Hall effect in Kagome lattices}
\author{H. Chen}
\author{H. Nassar}\email[Corresponding author: ]{nassarh@missouri.edu}
\affiliation{Department of Mechanical and Aerospace Engineering, University of Missouri, Columbia, MO 65211, USA}
\author{A. N. Norris}
\affiliation{Department of Mechanical and Aerospace Engineering, Rutgers University, Piscataway, NJ 08854-8058, USA}
\author{G.K. Hu}
\affiliation{School of Aerospace Engineering, Beijing Institute of Technology, Beijing 100081, China}
\author{G.L. Huang}\email[Corresponding author: ]{huangg@missouri.edu}
\affiliation{Department of Mechanical and Aerospace Engineering, University of Missouri, Columbia, MO 65211, USA}
\begin{abstract}
A Quantum Spin-Hall Insulator (QSHI) is implemented into a simple mass-spring Kagome lattice. The transition from the trivial state to the topological one is described by an invariant Chern number function of a contrast parameter. The band diagram and helical edge states characteristic of QSHI are obtained by a combination of numerical and analytical methods. In particular, these states are shown to be Stoneley wave solutions to a set of asymptotic continuous motion equations. Last, scatterless propagation of polarized topological edge waves around sharp corners is demonstrated and robustness is assessed through a parametric study.
\end{abstract}
\maketitle
\section{Introduction}
Similarities between the equations governing the physics of periodic lattices, whether acoustic, phononic, photonic or electronic, have allowed for establishing a sort of dictionary whereby wave phenomena occurring in one type of lattices can be translated, adapted and observed in other types of lattices as well.\cite{Lu2014,Yang2015,Huber2016} Recently, based on an analogy between the Hamiltonian of an electron in a crystal and the dynamical/stiffness matrix of a mechanical lattice, the concept of mechanical topological insulators has emerged.\cite{Susstrunk2015} Like their electronic predecessors,\cite{Hasan2010,Qi2011} they exhibit an insulating bulk and conducting polarized edge states immune to back-scattering by defects and corners. Edges of topological insulators thus constitute a novel class of superior waveguides with exceptionally robust transmission.

Phases of topological insulators are classified based on a quantized invariant, namely the Chern number, attached to a bulk bandgap. As long as the gap remains open, perturbing the constitutive and geometric parameters will have no influence on the qualities of the insulator which will therefore remain in the same phase with the same Chern number.\cite{Hasan2010,Qi2011} Conversely, changing the phase of an insulator requires closing the gap. Accordingly, lattices exhibiting Dirac cones, whereby two bands touch along a single point, are in a critical state: small perturbations that lift the degeneracy can toggle the lattice between a trivial phase with a zero Chern number and a topological phase with a non-zero Chern number. For the specific class of Quantum Spin-Hall Insulators (QSHI), the critical state exhibits in fact a double Dirac cone with a quadruple degeneracy. This suggests that designing a quantum-spin Hall insulator can be carried out in two steps: first, construct a lattice with a double Dirac cone. Second, classify perturbations that lift the degeneracy by their induced Chern numbers. It is noteworthy that QSHI-inducing perturbations must preserve time-reversal symmetry. In contrast, Quantum Hall Insulators (QHI) necessitate perturbations that break time-reversal symmetry and thus include non-time-invariant active components.\cite{Nash2015,Yang2015,Khanikaev2015,Fleury2016,Nassar2018} Other topological phases such as the quantum valley-Hall phase are based on breaking or keeping other spatial symmetries such as inversion, $C_3$ or $C_6$ symmetry.\cite{Lu2016a,Lu2016,Liu2017,Pal2017,Vila2017,Ni2017,Chen2018}

Starting with a lattice exhibiting two Dirac cones, a double Dirac cone can be obtained by one of two ways. If the two initial cones have different frequencies, then the geometric and constitutive parameters of the lattice need to be optimized so as to bring the two cones arbitrarily close to one another.\cite{Mousavi2015,Miniaci2017} Alternatively, if the two initial cones have the same frequency, Brillouin zone folding techniques can be used to superimpose the two cones.\cite{Zhang2017,Yves2017a,Yves2017,Deng2017,Xia2017,Chaunsali2018} In comparison, the second technique is simpler to implement, especially in hexagonal lattices that, due to symmetry, exhibit by default two Dirac cones at $K$ and $K'$ points which, by folding, end up laying on top of one another at the $\Gamma$ point.

In this letter, a QSHI is implemented in a hexagonal Kagome lattice using the Brillouin zone folding technique. Note that previous implementations of the QSHI either used impractical bi-layered unit cell designs,\cite{Susstrunk2015,Pal2016} or full plate models with coupled in-plane and out-of-plane modes.\cite{Mousavi2015,Miniaci2017,Chaunsali2018} The virtue of the present design resides in its simplicity: it is single-layered, has a limited number of Degrees Of Freedom (DOF) per unit cell and is genuinely two-dimensional with no out-of-plane components. The QSHI, in terms of its bulk and edge spectra, associated polarized edge states, their shapes and decay speeds, is characterized using a continuum asymptotic homogenized model further consolidated by numerical simulations. Of particular interest, is the demonstrated ability of the edge of the QSHI to propagate scatterless edge states around corners and defects. Finally, we assess the robustness of the QSHI and analyze the existence of a localization-transmission trade-off.
\begin{figure}[H]
\centering
\includegraphics[width=\linewidth]{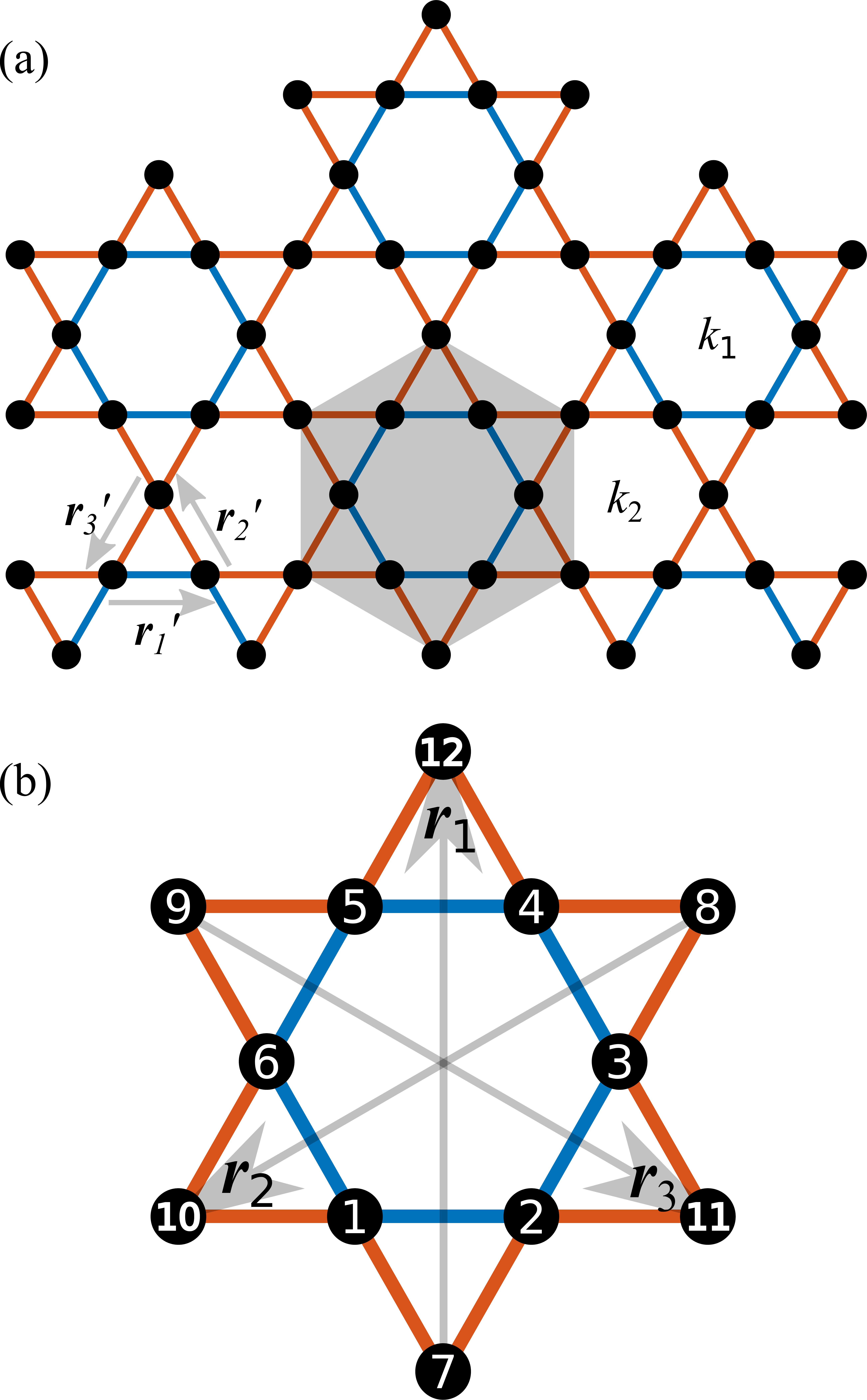}
\caption{The perturbed Kagome lattice (a) and a magnified view of its unit cell (b).}
\label{fig:lattice-SM}
\end{figure}
\section{Complete model}
Consider the Kagome lattice annotated on Figure~\ref{fig:lattice-SM}a. It is periodic invariant by translation along vectors $\t r_1$, $\t r_2$, $\t r_3$ and any integer combination thereof. Its dispersion diagram, eigenfrequencies and natural modes can be derived by analyzing the motion of one unit cell (Figure~\ref{fig:lattice-SM}b) under Floquet-Bloch boundary conditions. Thus, let $\t q$ be a wavenumber, and $\t u_j$ be the displacement vector of mass number $j$, then the boundary conditions are
\[
\t u_{12} = Q_1 \t u_7,\quad
\t u_{10} = Q_2 \t u_8,\quad
\t u_{11} = Q_3 \t u_9,
\]
with $Q_j = e^{i\av{\t q,\t r_j}}$. Accordingly, the motion equation of mass $1$ for instance can be written as
\begin{multline}
-\omega^2 m \t u_1 = k_1\av{\t u_2-\t u_1,\t r_1'}\t r_1'
+k_1\av{\t u_6-\t u_1,\t r_2'}\t r_2'\\+
k_2\av{Q_2\t u_8-\t u_1,\t r_1'}\t r_1'+
k_2\av{\t u_7-\t u_1,\t r_2'}\t r_2',
\end{multline}
where $k_1$ and $k_2$ are spring constants and $m$ is mass. In matrix form, this reads
\begin{multline}\label{eq:u1}
-\omega^2 m \t u_1 = 
(k_1+k_2)(\t r_{11}'+\t r_{22}')\t u_1 
-k_1 \t r_{11}'\t u_2\\
-k_1\t r_{22}'\t u_6
-k_2\t r_{22}'\t u_7
-k_2Q_2\t r_{11}'\t u_8
\end{multline}
or more compactly
\[
-\omega^2 m \begin{bmatrix} \t u_1 \end{bmatrix} = -L_1 \begin{bmatrix}
\t u_1 \\ \vdots \\ \t u_9
\end{bmatrix}
\]
where $L_1$ is a $2\times 18$ matrix whose entries can be deduced from~\refeq{eq:u1} with
\[
\t r_1 = \begin{bmatrix}
0 \\ 1
\end{bmatrix}, \quad
\t r_2 = \begin{bmatrix}
-\sqrt 3 /2 \\ 1/2
\end{bmatrix}, \quad
\t r_3 = \begin{bmatrix}
\sqrt 3 / 2 \\ -1/2
\end{bmatrix},
\]
and
\[
\t r_1' = \begin{bmatrix}
1 \\ 0
\end{bmatrix}, \quad
\t r_2' = \begin{bmatrix}
-1/2 \\ \sqrt 3 /2
\end{bmatrix}, \quad
\t r_3' = \begin{bmatrix}
-1/2 \\ -\sqrt 3 / 2
\end{bmatrix},
\]
and finally
\[
\t r_{jj}' = \t r_j' \t r_j'^T.
\]
By carrying similar calculations for the eight other masses, the motion equation can be put in the form
\[
-\omega^2 m
\begin{bmatrix}
\t u_1 \\ \vdots \\ \t u_9
\end{bmatrix}
=
-
\begin{bmatrix}
L_1 \\ \vdots \\ L_9
\end{bmatrix}
\begin{bmatrix}
\t u_1 \\ \vdots \\ \t u_9
\end{bmatrix}
\]
or, more symbolically,
\[
\omega^2 m \Phi = H \Phi.
\]
In a non-dimensional form, this becomes
\[
\Omega^2 \Phi = \hat H \Phi,\quad \hat H = H/k, \quad \Omega^2 = \omega^2 m/k,
\]
with $k=(k_1+k_2)/2$. Last, the associated dispersion diagram is deduced from the zero-determinant condition
\[
\det(\hat H-\Omega^2 I) = 0
\]
where $I$ is the $18\times 18$ identity matrix.

\begin{figure}[H]
\centering
\includegraphics[width=\linewidth]{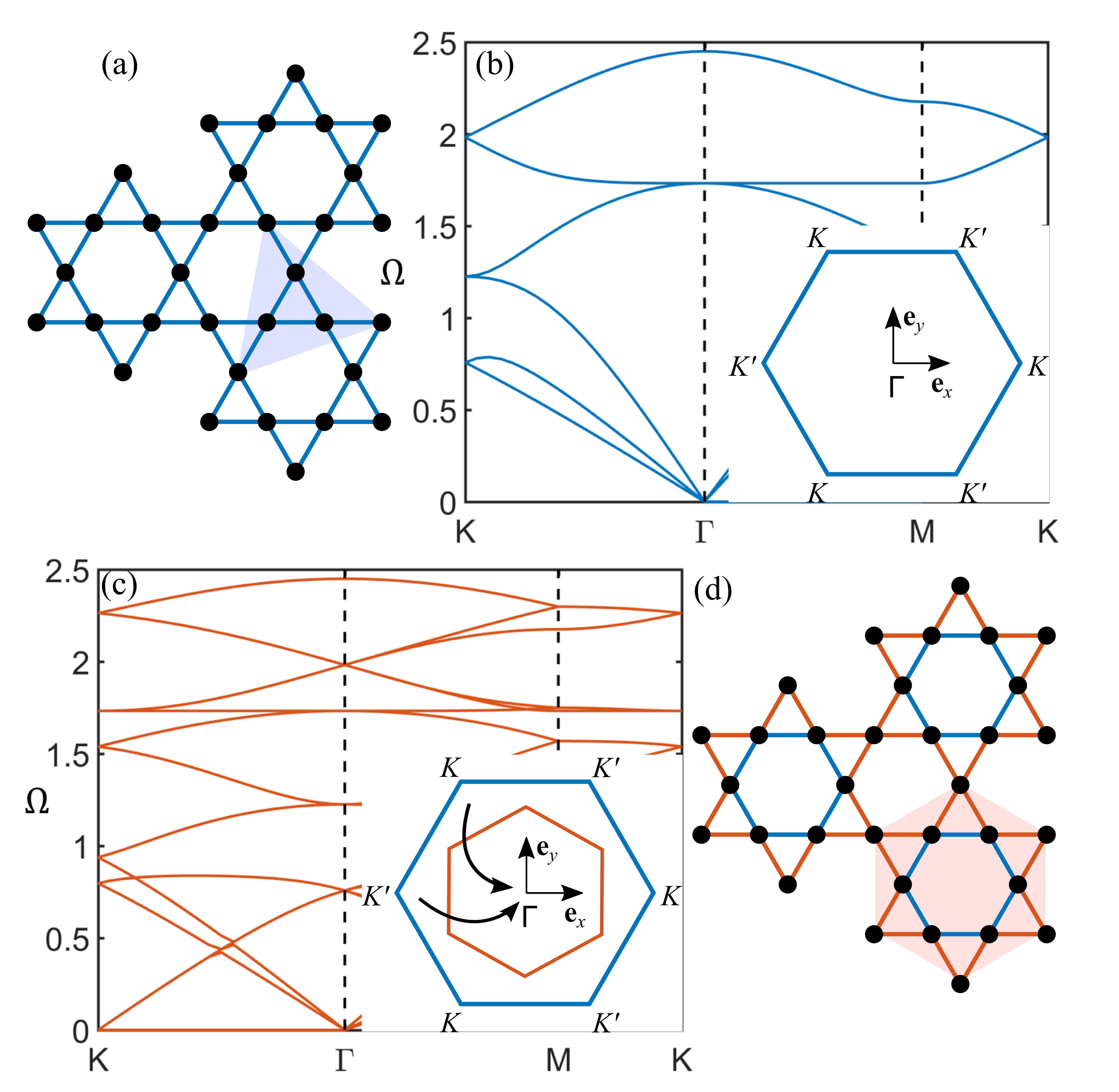}
\caption{The Kagome lattice: (a) homogeneous configuration; (b) its dispersion diagram featuring two Dirac cones between the fifth and sixth bands at $K$ and $K'$; (c) folded dispersion diagram featuring a double Dirac cone at the $\Gamma$ point; (d) the corresponding perturbed configuration. Note that diagram (c) correspond to the lattice in (d) only in the limit $k_1=k_2$.}
\label{fig:lattice}
\end{figure}

The host Kagome spring-mass lattice is illustrated in Figure~\ref{fig:lattice}a and initially has three masses per unit cell totaling six DOFs. It further exhibits two inequivalent Dirac cones with the same frequency at the corners $K$ and $K'$ of the Brillouin zone shown on Figure~\ref{fig:lattice}b. By folding the Brillouin zone along the bisectors of $\Gamma K$ and $\Gamma K'$, these two Dirac cones will form together a double Dirac cone at the $\Gamma$ point. The resulting folded diagram along with the folding motion are shown on Figure~\ref{fig:lattice}c. The folding can be induced by periodically perturbing the constants of the springs with nearly twice the original space period. Here, a star-shaped perturbation is adopted as depicted on Figure~\ref{fig:lattice}d. A unit cell now has nine masses for a total of 18 DOFs.

Let $\beta=(k_1-k_2)/(k_1+k_2)$ denote the relative contrast between spring constants. For $\beta=0$ and $k_1=k_2$ a reduced unit cell with only three masses and six DOFs can be chosen as in Figure~\ref{fig:lattice}a. The resulting dispersion diagram is plotted on Figure~\ref{fig:lattice}b and exhibits two inequivalent Dirac cones with the same normalized frequency $\Omega_o=\sqrt{3(3+\sqrt{5})}/2$ at the corners $K$ and $K'$ of the Brillouin zone. By folding the Brillouin zone along the bisectors of $\Gamma K$ and $\Gamma K'$, these two Dirac cones will form together a double Dirac cone at the $\Gamma$ point. The resulting folded diagram along with the folding motion are shown on Figure~\ref{fig:lattice}c. The folding can be induced by periodically perturbing the constants $k_1$ and $k_2$ of the springs so that the choice of the unit cell highlighted on Figure~\ref{fig:lattice}d becomes necessary.

\begin{figure*}[ht!]
\centering
\includegraphics[width=\linewidth]{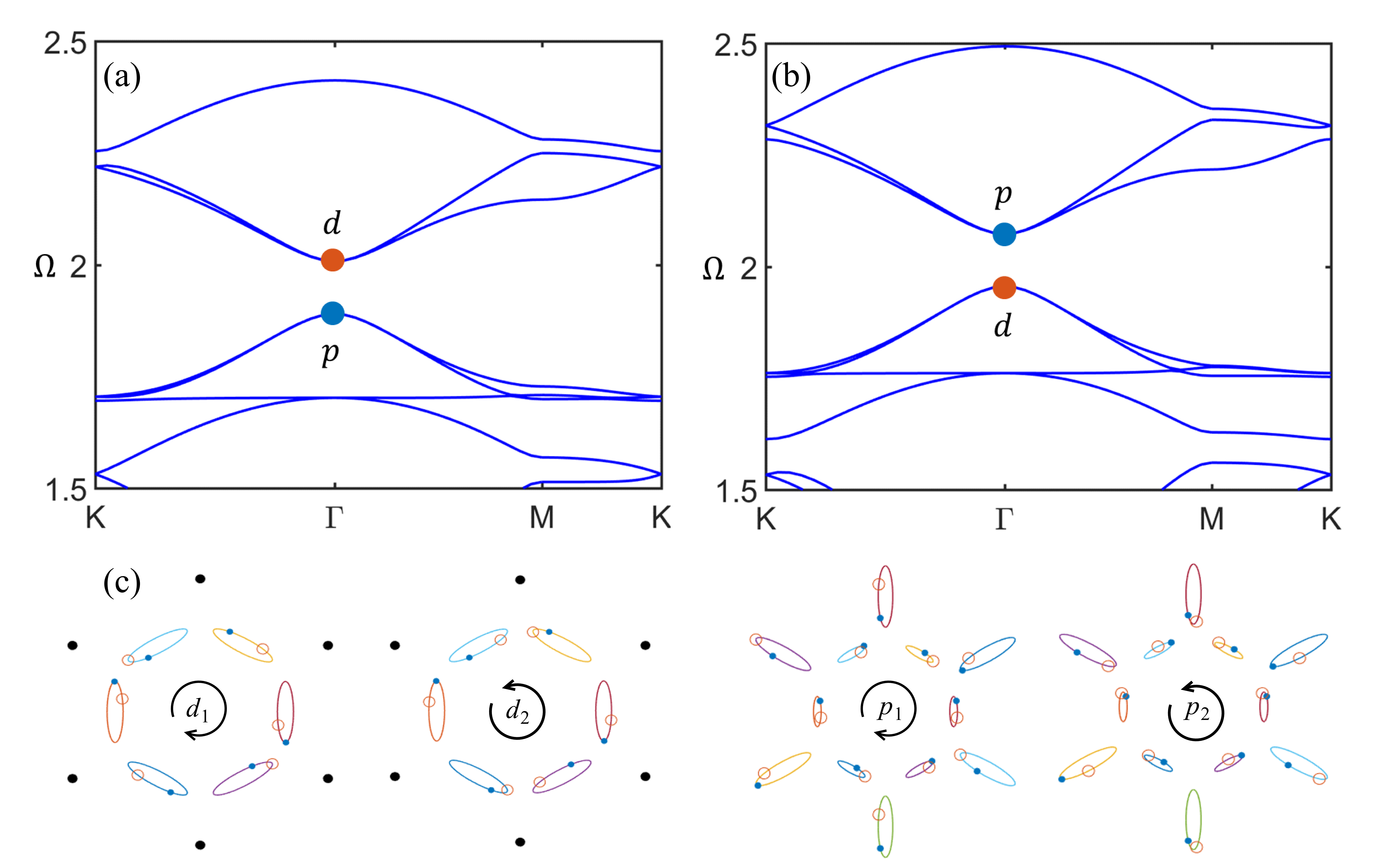}
\caption{Band inversion: dispersion diagrams for $\beta=0.1$ (a) and $\beta=-0.1$ (b). As $\beta\neq 0$, the four-fold degeneracy breaks into two two-fold degeneracies. The mass trajectories of the degenerate eigenstates are illustrated in (c) along with their polarizations: a solid blue dot corresponds to the initial position of the mass and an orange circle corresponds to the position at a later small time. Black dots correspond to non-moving masses.}
\label{fig:mode1}
\end{figure*}

The folded dispersion diagram of Figure~\ref{fig:lattice}c corresponds to the case where $\beta$ is infinitely close to $0$. The four-fold degeneracy of the double cone is partially lifted for non-zero $\beta$ values and breaks into two two-fold degeneracies with frequencies $\Omega_p$ for modes $p_{1,2}$ and $\Omega_d$ for modes $d_{1,2}$; see Figure~\ref{fig:mode1}a,b. Modes $d_{1,2}$ have zero displacements at the boundaries of the unit cell and have pairs of diametrically opposed masses moving in phase. Conversely, modes $p_{1,2}$ have maximum displacements at the boundaries of the unit cell and have pairs of diametrically opposed masses moving in opposition of phase. Last, for $\beta>0$, modes $p_{1,2}$ have a lower frequency $\Omega_p<\Omega_d$ than modes $d_{1,2}$ as they solicit mainly the softer of the two springs with constant $k_2$. As $\beta$ decreases however and crosses $0$, frequencies and modes exchange places as the softer and stiffer springs exchange places. This band inversion phenomenon illustrated on Figure~\ref{fig:mode1} is symptomatic of a topological phase transition occurring between a trivial and a non-trivial state and is investigated next in terms of an asymptotic model.
\section{Asymptotic model}
When $\beta$ is close to $0$, $q$ to $0$ and $\Omega$ to the double Dirac cone frequency $\Omega_o$, the displacement field $\Phi$ approaches the space spanned by the degenerate eigenmodes $(p_1,p_2,d_1,d_2)$ at $(q=0,\Omega=\Omega_o)$. Accordingly, there exists four complex numbers interpreted as generalized coordinates $(\xi_1,\xi_2,\zeta_1,\zeta_2)$ such that
\[
\Phi = \xi_1 p_1 + \xi_2 p_2 + \zeta_1 d_1 + \zeta_2 d_2 + \delta\Phi, \quad
\norm{\delta\Phi}\ll \norm{\Phi},
\]
where $\delta\Phi$ gathers first order corrections to the displacements of the mode $\Phi$. Introducing the reduced coordinate vector $\phi$ and the projector matrix $P$ given by
\[
\phi = \begin{bmatrix}
\xi_1 \\ \zeta_1 \\ \xi_2 \\ \zeta_2
\end{bmatrix},\quad
P = \begin{bmatrix}
p_1 & d_1 & p_2 & d_2
\end{bmatrix}
\]
one can write
\[
\Phi = P\phi + \delta\Phi.
\]
In a similar manner, the dynamical matrix $\hat H$ can be Taylor expanded as
\[
\hat H = \hat H_o + \delta \hat H,
\]
with $\hat H_o=\hat H(q=0,\beta=0)$ being the leading order dynamical matrix at zero contrast and at the double Dirac cone and
\[
\delta \hat H = \beta\partial_\beta \hat H + q_x\partial_{q_x} \hat H + q_y\partial_{q_y} \hat H
\]
being its first order correction composed of two terms: one due to the presence of a small non-zero contrast $\beta$, the other due to a small non-zero wavenumber $\t q=(q_x,q_y)$.

Substituting all expansions into the motion equation leads to
\[
(\hat H_o+\delta \hat H)(P\phi+\delta\Phi) = (\Omega_o^2+\delta\Omega^2)(P\phi+\delta\Phi)
\]
which simplifies into
\[
\delta \hat H P\phi+ \hat H_o\delta\Phi = \Omega_o^2\delta\Phi + \delta\Omega^2 P\phi.
\]
Applying the projector $P$ to the above equation finally entails
\[\label{eq:AME}
P^\dagger\delta\hat H P\phi =  \delta\Omega^2 P^\dagger P\phi,
\]
where $P^\dagger P$ is the $4\times 4$ identity matrix by orthogonality of the eigenmodes and the effective dynamical matrix $\delta h \equiv P^\dagger\delta \hat H P$ is evaluated to be
\[
\delta h
=
\begin{bmatrix}
\delta h^+ & 0 \\
0 & \delta h^-
\end{bmatrix},
\]
with
\[
\delta h^+ = 
\begin{bmatrix}
- a\beta & b(\delta q_x + i\delta q_y) \\ * & \beta
\end{bmatrix},
\]
and
\[
\delta h^- = 
\begin{bmatrix}
- a\beta & -b^*(\delta q_x - i\delta q_y) \\ * & \beta
\end{bmatrix},
\]
Here, $a\approx 3.7$ and $b\approx 0.17 - 0.32i$ are non-dimensional numerical factors function of the geometry of the lattice.

Analytical approximations to the dispersion diagram near the double Dirac cone can be derived thanks to the zero-determinant condition now written in terms of $\delta h$ instead of $H$ (Figure~\ref{fig:mode1}). It reads
\[
2\Omega_o(\Omega - \Omega_o) = \frac{1-a}{2}\beta \pm\sqrt{\left(\frac{1+a}{2}\right)^2\beta^2 + \lvert b\rvert^2(q_x^2 + q_y^2)}.
\]
and describes a cone in the case $\beta=0$. For $\beta\neq 0$, the eigenfrequencies of the modes $d_{1,2}$ and $p_{1,2}$ can be deduced by letting $q_x=q_y=0$:
\[
\Omega_d = \Omega_o+\frac{\beta}{2\Omega_o}, \quad
\Omega_p = \Omega_o-\frac{a\beta}{2\Omega_o}.
\]

The topological invariant, namely the Chern number, can be calculated according to\cite{Mousavi2015}
\[
c = \frac{1}{2\pi}\iint_{BZ} F(q)\dd q_x\dd q_y,\quad F(q) = -i\nabla_q \times \av{\Psi,\nabla_q \Psi},
\]
for $\beta\neq 0$. Therein, $\Psi$ is the eigenmode of matrix $\delta h^+$ with the highest eigenvalue. We find $c=0$ for $\beta > 0$ and $c=\pm 1$ for $\beta<0$ confirming indeed the occurrence of a topological phase transition from a trivial state $(\beta>0)$ to a topological one $(\beta<0)$ as $\beta$ crosses $0$ and the bands are inverted.
\section{Edge and interface states}
\subsection{Numerical analysis}
\begin{figure}[ht!]
\centering
\includegraphics[width=\linewidth]{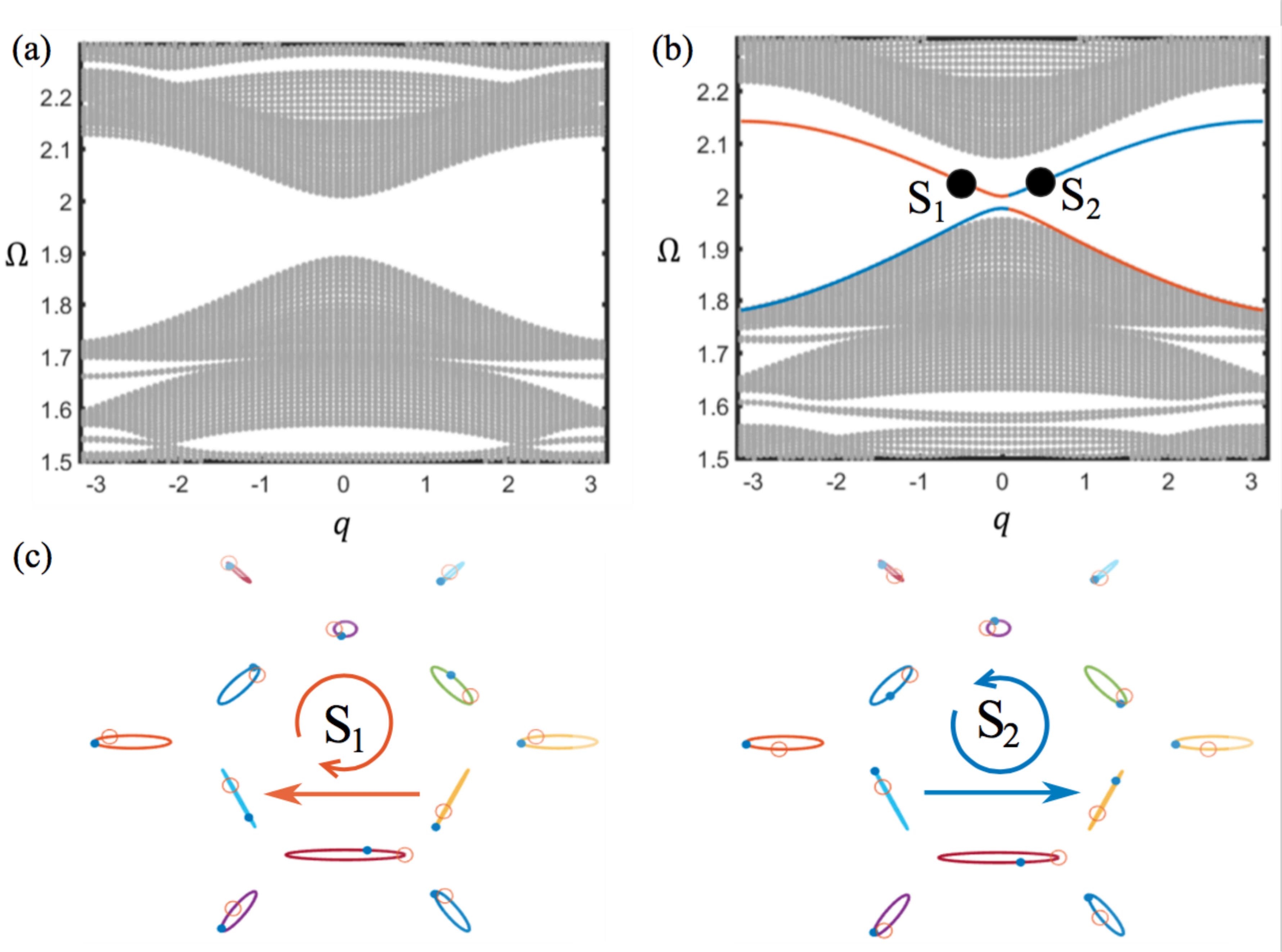}
\caption{Bulk and edge spectra of a finite sample of 20 unit cells under free (resp., periodic) boundary conditions in the $y$-direction (resp., $x$-direction) for $\beta=0.1$ (a) and $\beta=-0.1$ (b). In the latter case, edge modes exist and the corresponding mass trajectories are illustrated on (c) for $q=\pm 0.5$.}
\label{fig:edge}
\end{figure}

By the bulk-edge correspondence principle, the total bulk bandgap for a topological lattice, i.e., with $\beta<0$, will host a pair of helical states localized at edges with opposite polarizations and opposite directions of propagation. To confirm this behavior, the bulk and edge spectra for $\beta$ positive and negative are plotted on Figure~\ref{fig:edge} for a finite slab under free boundary conditions in the $y$-direction and under Floquet-Bloch boundary conditions in the $x$-direction. It is then seen that a pair of edge states populate the bulk bandgap in the case $\beta<0$ confirming the predictions of the bulk-edge correspondence principle and that the phase with $\beta<0$ (resp., $\beta>0$) is topological (resp., trivial). These edge states are helical or elliptically polarized with opposite polarizations and opposite propagation directions. It is worth mentioning that these edge states are not completely gapless as would be expected in a genuine QSHI. Here, due to the breaking of $C_6$ symmetry near boundaries, a small edge bangap appears and is expected to be as small as $\beta$ is.

\begin{figure}[H]
\centering
\includegraphics[width=\linewidth]{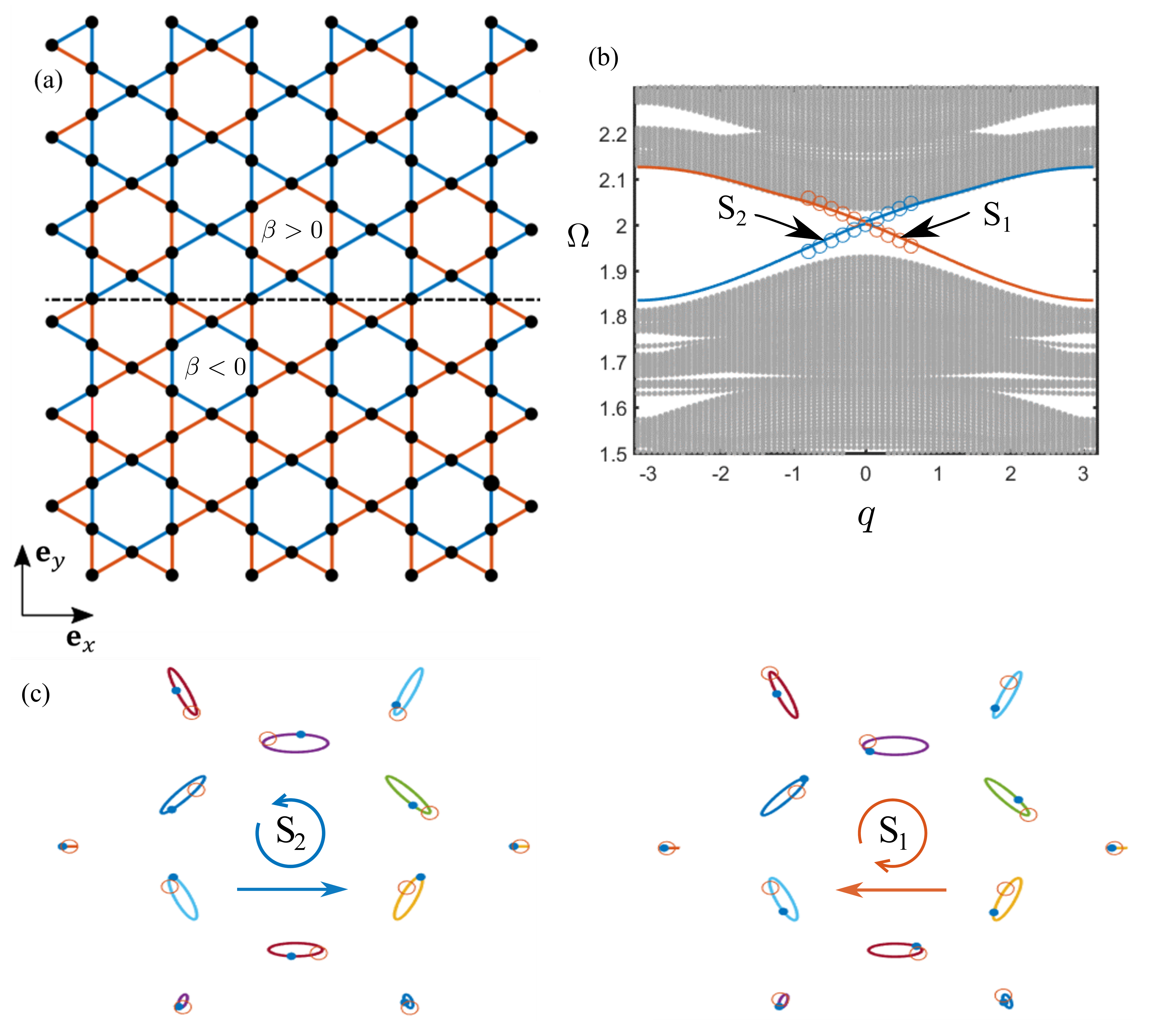}
\caption{Bulk and edge spectra of a finite sample of 40 unit cells under free (resp., periodic) boundary conditions in the $y$-direction (resp., $x$-direction) with $\beta=0.2$ for $y>0$ and $\beta=-0.2$ for $y<0$: (a) geometry; (b) diagram: continuous lines correspond to the numerical result and circles correspond to the asymptotic result (equations~\refeq{eq:range} and~\refeq{eq:disp}) ; (c) mass trajectories of interface states for $q=\pm 0.5$.}
\label{fig:interface}
\end{figure}

In a similar fashion, we calculated the edge states localized at the interface between two lattices with opposite contrast (Figure~\ref{fig:interface}a). The sample is composed of $40$ unit cells with periodic boundary conditions in the $x$-direction and free boundary conditions in the $y$-direction. The first half of the unit cells below the interface $y=0$ has a negative contrast whereas the second half, above $y=0$, has a positive contrast. On Figure~\ref{fig:interface}b, the bulk and edge spectra are illustrated. The mass trajectories corresponding to these interface waves are plotted on Figure~\ref{fig:interface}c and again are oppositely polarized.
\subsection{Continuum analysis}
Using the asymptotic model of the previous section, the calculated interface modes can be interpreted as Stoneley waves and can be derived as solutions to~\refeq{eq:AME} recast in differential form. Thus,
let $\beta$ be a function of $y$ independent of $x$. By symbolically mapping $iq_y \mapsto \partial_y$, we obtain a pair of continuum motion equations
\[
\begin{split}
-a\beta(y) \xi_1(y) + b(q_x + \partial_y)\zeta_1(y) &= \delta\Omega^2 \xi_1,\\
b^*(q_x - \partial_y)\xi_1(y) + \beta(y) \zeta_1(y)  &= \delta\Omega^2 \zeta_1
\end{split}
\]
governing $(\xi_1,\zeta_1)$ and a similar pair of equations for $(\xi_2,\zeta_2)$. Now consider an interface $y=0$ separating between two zones, one with $\beta = \beta_o > 0$, say for $y>0$, and the other with $\beta=-\beta_o$ for $y<0$. Then, a Stoneley wave solution localized at $y=0$ can be derived. It is given by
\[
\xi_1 = \alpha_{\pm} \zeta_1 = \alpha_{\pm} A_{\pm} \exp\left(Q_{\pm}y\right)
\]
where the sign $+$, resp. $-$, is adopted for $y>0$, resp. $y<0$. This solution must satisfy four conditions, two ensuring exponential decay at $y=\pm\infty$, and two ensuring continuity across the interface $y=0$. These can be satisfied as long as:
\[
A_+ = A_- \equiv A,\quad \alpha_+=\alpha_-\equiv \alpha, \quad Q_+<0,\quad Q_- > 0.
\]
Injecting the expression of the Stoneley wave in the continuum motion equations, the decay conditions imply 
\[\label{eq:range}
Q_{\pm}^2 = q_x^2-\frac{(\delta\Omega^2\pm a\beta_o)(\delta\Omega^2\mp\beta_o)}{\lvert b\rvert^2},
\]
whereas the continuity conditions reduce to
\[\label{eq:disp}
(a\beta_o-\delta\Omega^2)Q_+ + (a\beta_o+\delta\Omega^2)Q_- = -2a\beta_o q_x.
\]
Consequently, equation~\refeq{eq:range} provides the range of existence of Stoneley waves whereas equation~\refeq{eq:disp} is a relationship between $\delta\Omega^2$ and $q_x$ and corresponds to their dispersion relation.
\begin{figure}
\centering
\includegraphics[width=\linewidth]{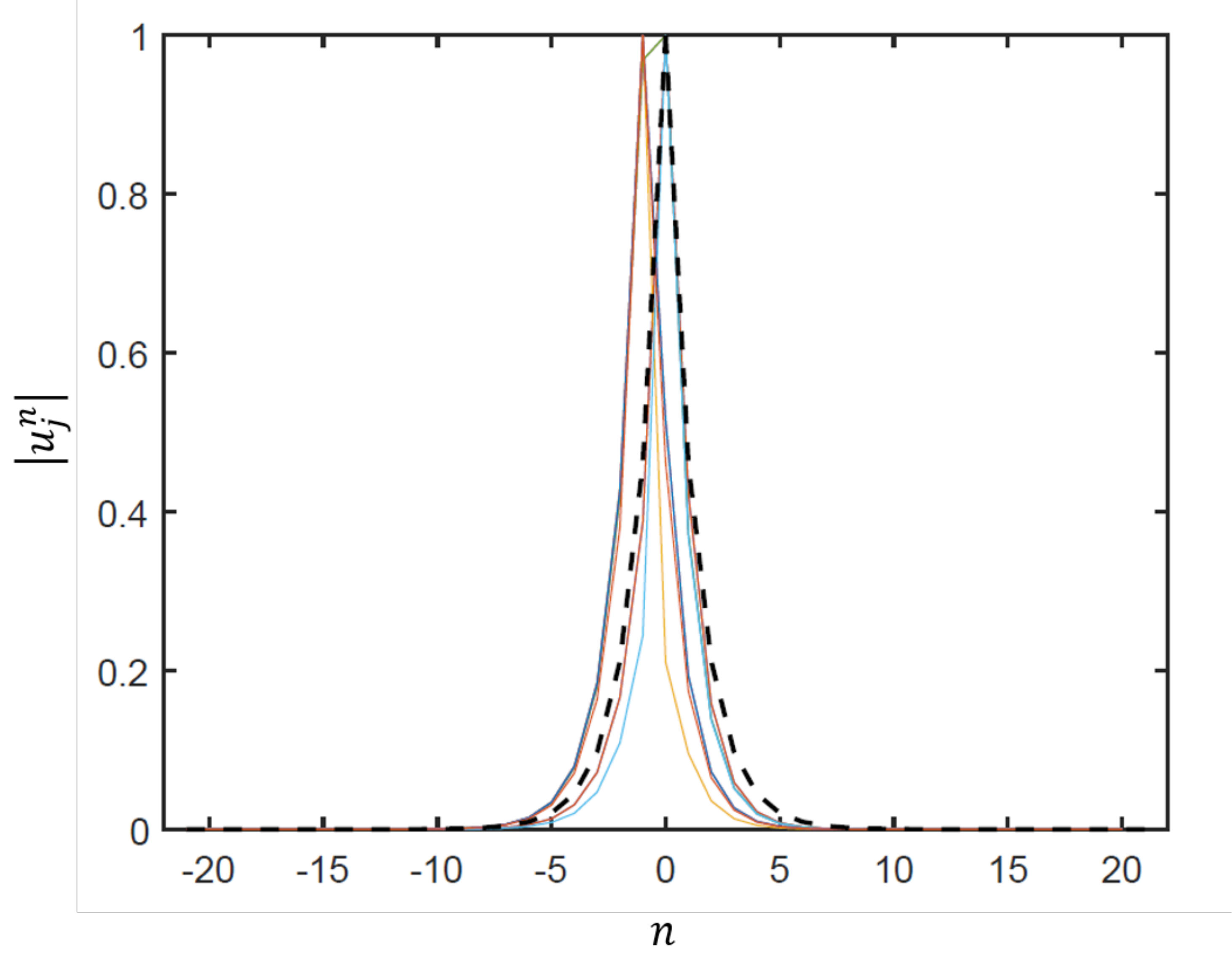}
\caption{Decay profiles of the Stoneley wave at $q_x=0$ calculated numerically (solid) and asymptotically (dashed).}
\label{fig:StoneProfile}
\end{figure}

Finally, the full Stoneley wave reads
\[\label{eq:S1}
\Phi = A(\alpha p_1+ d_1)\exp\left(Q_\pm y\right)\exp[i(q_x x-\Omega t)]
\]
where $A$ is a complex amplitude. We highlight that another Stoneley wave with an opposite group velocity exists based on the modes $p_2$ and $d_2$. It can be deduced from the one exhibited above by time-reversal symmetry $t\mapsto -t$. Alternatively, it is given by
\[\label{eq:S2}
\Phi = A(\alpha p_2+ d_2)\exp\left(Q_\pm y\right)\exp[i(-q_x x-\Omega t)].
\]
The dispersion relation of the Stoneley waves is plotted on Figure~\ref{fig:interface}b based on equations~\refeq{eq:range} and~\refeq{eq:disp} and their decay profile is depicted on Figure~\ref{fig:StoneProfile}. Both figures show good agreement between asymptotic and full models.
\begin{figure}[H]
\centering
\includegraphics[width=\linewidth]{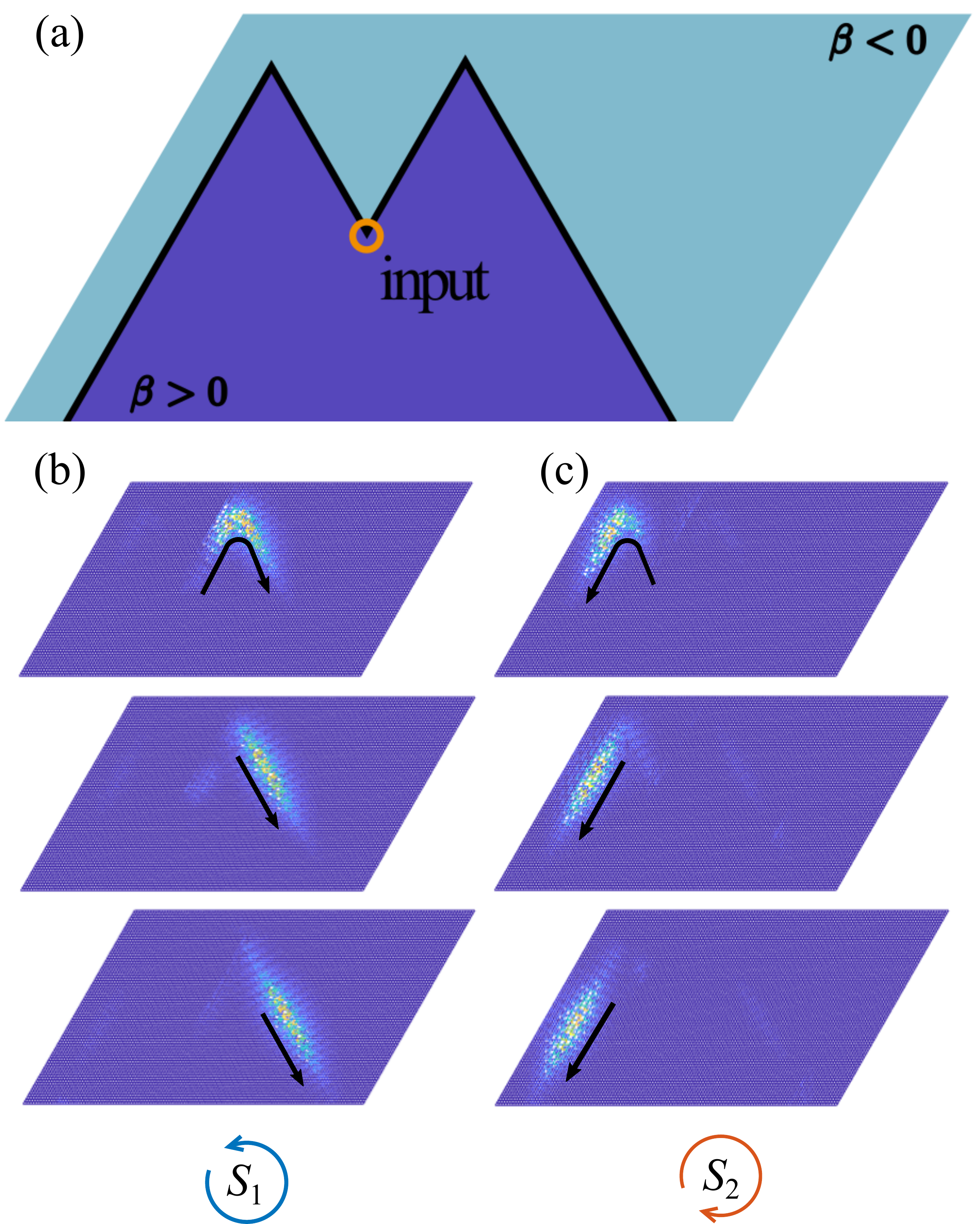}
\caption{Immunity to backscattering at corners: (a) geometry; (b) snapshots of the displacement amplitude as a color map for a positively polarized excited wave at $t=200, 250, 300$ respectively ; (c) same as (b) for a negatively polarized excited wave.}
\label{fig:transient}
\end{figure}
\begin{figure*}
\centering
\includegraphics[width=\linewidth]{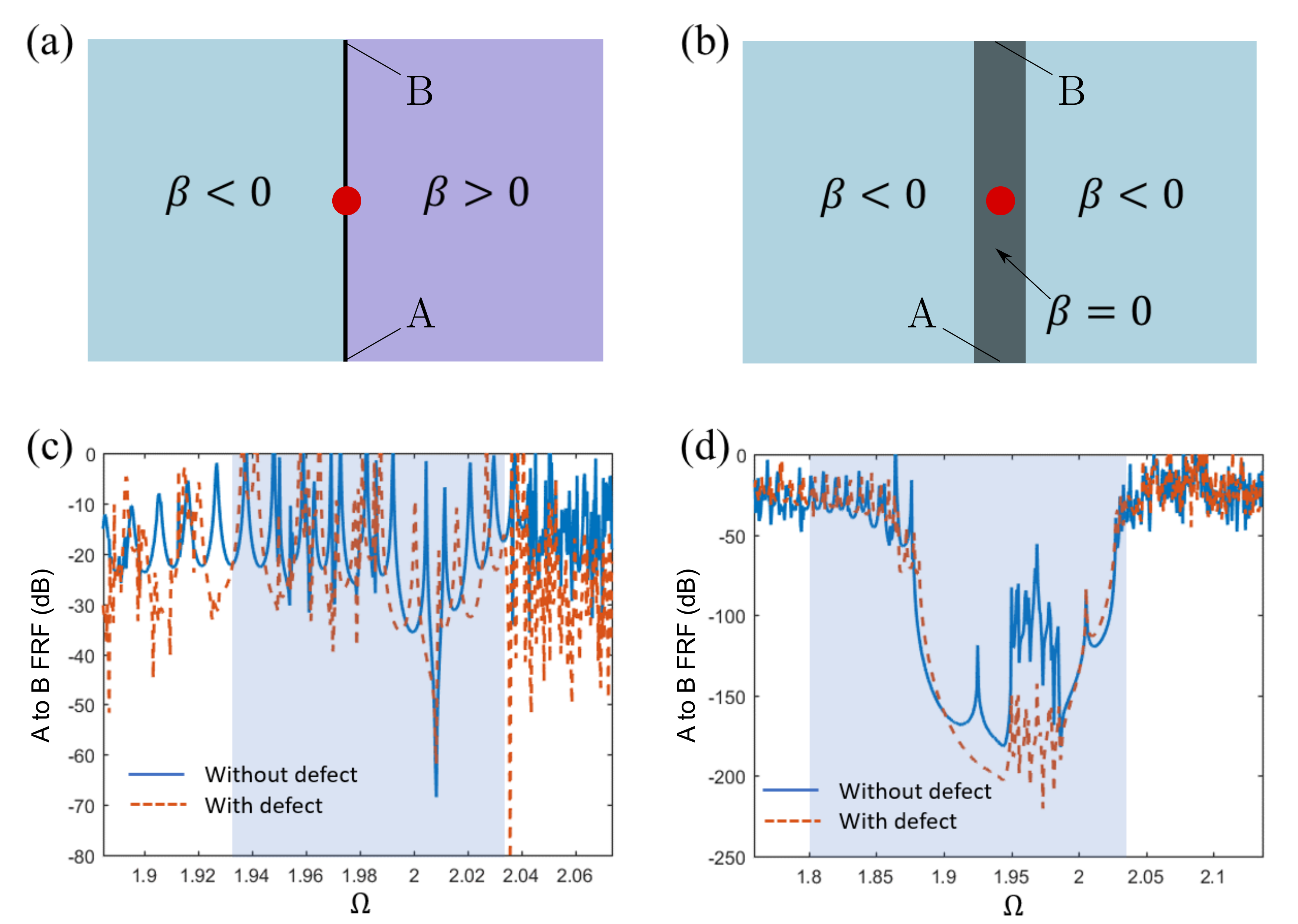}
\caption{\label{fig:r1}Assessing robustness of a topological. (a) A topological waveguide. (b) A trivial waveguide. (b,c) Frequency response functions in function of frequency with and without defects: displacement is imposed at A and collected at B and the ratio is ploted on a log scale; the bulk bandgap is highlighted; and the defect, when present, consists in removing half a unit cell where the red dot is located on (a,b). Simulations carried with $\beta=\pm 0.2$}
\end{figure*}
\section{Transient analysis}
The interface Stoneley waves given in equations~\ref{eq:S1} and~\ref{eq:S2} are practically uncoupled: defects that do not break time-reversal symmetry cannot backscatter one mode into the other. To illustrate that fact, transient numerical simulations of a signal propagated along an interface featuring a sharp corner are carried. The interface is M-shaped and separates one trivial and one topological domain as shown in Figure~\ref{fig:transient}a. A tone-burst loading of central frequency $\Omega_o$ is applied to the middle tip and is calculated so as to excite the positively polarized mode only. The loading thus excites a single wave going right shown on Figure~\ref{fig:transient}b. At the next tip, the propagated wave makes the turn following the interface with zero backscattering. On Figure~\ref{fig:transient}c, the negatively polarized wave is excited and undertakes a similar scatterless path. In both cases, the tone-burst loading's band lies mostly within the bulk bandgap so as not to generate any bulk waves. Simulations were carried under free boundary conditions using the discrete spring-mass elements of the commercial software ANSYS.
\section{Assessing robustness}
It is of interest to quantify the extent of the topological protection presented above. Thus, we calculate the end-to-end frequency response function of a topological waveguide (Figure~\ref{fig:r1}a) with and without a small defect consisting of removing one half of the unit cell located midway at the interface (red dot). For reference, the performance of the topological waveguide is compared to that of a trivial waveguide designed by sandwiching a thin slice of a uniform Kagome lattice ($\beta=0$) between two gapped but \emph{topologically equivalent} lattices (e.g., $\beta<0$ to both sides; see Figure~\ref{fig:r1}b). The results are plotted on Figure~\ref{fig:r1}c,d. Within the bulk bandgap (highlighted zone, it is observed that the topological waveguide with and without defect performs mostly the same with response levels of the same order of magnitude as outside the bulk bandgap. In contrast, the response levels of the trivial waveguide significantly drop around the center of the gap without the defect, and even more so in the presence of the defect. It is clear then that the topological waveguide is immune to such small defects whereas the trivial one is extremely sensitive to their presence.
\begin{figure}[ht!]
\centering
\includegraphics[width=\linewidth]{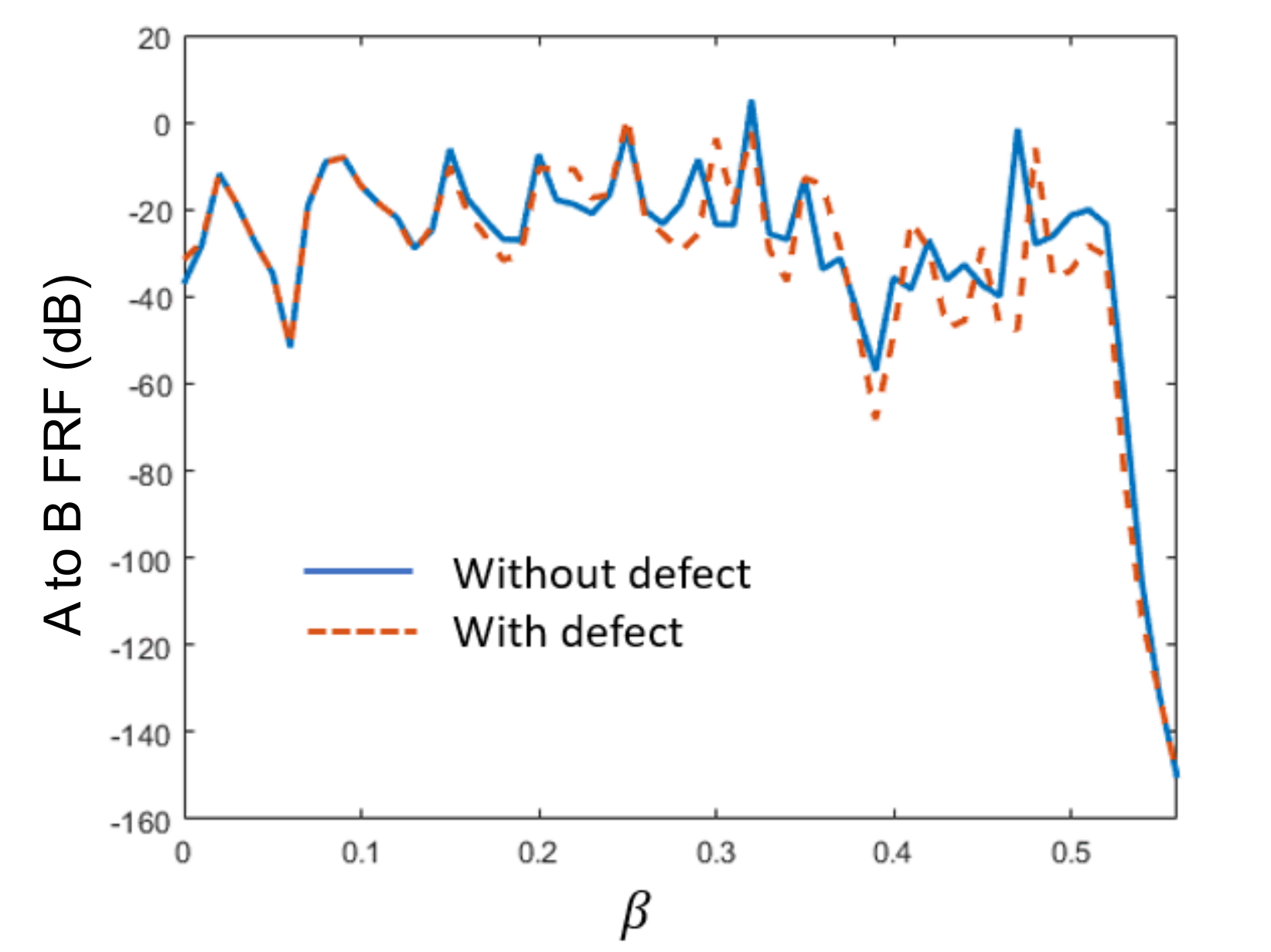}
\caption{Frequency response function of a topological waveguide with and without defect at the Dirac frequency for varying contrast $\beta$.}
\label{fig:r2}
\end{figure}

However, it should not be concluded that topological protection is absolute. As a matter of fact, the performance of the topological waveguide can deteriorate while remaining insensitive to defects. Consider for instance the plot of Figure~\ref{fig:r2}: it shows the end-to-end frequency response function of the topological waveguide of Figure~\ref{fig:r1}a for varying $\beta$. It is seen then that, although the response is little-to-no sensitive to the presence of the defect, as $\beta$ increases beyond $0.5$, the topological waveguide fails in fulfilling its duty in guiding signals as its response drops to near zero levels. As a matter of fact, as $\beta$ increases the edge bandgap, previously observed on Figure~\ref{fig:edge}b, is enlarged and ultimately a total, bulk and edge, bandgap appears and forbids all signals to be transmitted. There is therefore a seemingly unsurmoutable trade-off between transmission levels on one hand and degree of localization on the other hand. As a matter of fact, the transmitted edge states are as localized near the interface as the bulk bandgap is wide and both are proportional to $\beta$ at leading order. However, as $\beta$ is increased to achieve more localized states, the edge bandgap widens and the transmission levels drop drastically. The parametric study shown here suggests that the optimal combination of localization and transmission is achieved around $\beta=0.5$, i.e., maximum contrast value for which transmission is unharmed.
\section{Conclusion}
As demonstrated, a QSHI can be implemented for classical mechanical waves in a fairly simple system such as the Kagome lattice. The coupling induced by a periodic perturbation to the spring constants is enough to make appear electronic or quantum mechanical features such as pseudo-spins and helically polarized states. The carried transient numerical simulations show that these states are uncoupled and can be used to transmit signals around specific defects, geometric or constitutive, without backscattering, i.e., with no loss of power. The edge of a QSHI, in a precise bandwidth, therefore acts as a robust waveguide with consistent high transmittance close to unity.
\begin{acknowledgments}
This work is supported by the Air Force Office of Scientific Research under Grants No. AF 9550-15-1-0016 and AF 9550-18-0096 with Program Manager Dr. Byung-Lip (Les) Lee, the NSF EFRI under award No. 1641078 and the Army Research office under Grant No. W911NF-18-1-0031 with Program Manager Dr. David M. Stepp. The 111 project (B16003) is also acknowledged.
\end{acknowledgments}
%
\end{document}